\documentclass[3p,times,twocolumn]{elsarticle}
 \biboptions{comma,sort&compress}
 
\usepackage{graphicx}
\usepackage{here}
\usepackage{ecrc}

\usepackage{slashed}
\usepackage{url}
\usepackage{bm}

\newcommand{\eqnsgn}[1]{\!\!\!\!& #1 &\!\!\!\!}

\volume{00}

\firstpage{1}

\journalname{Nuclear and Particle Physics Proceedings}

\runauth{}

\jid{nppp}

\jnltitlelogo{Nuclear and Particle Physics Proceedings}

\usepackage{amssymb}
\usepackage[figuresright]{rotating}

\begin{document}

\begin{frontmatter}

%%
%%%%%%%%%%%%%%%%%%%%%%%%%%%%%%%%%%%%%%%%%%%%%%%%%
\title{ Finite temperature corrections to a NLO Nambu-Jona-Lasinio model
 $^*$}
 % \corref{cor0}}
 \cortext[cor0]{Talk given at 19th International Conference in Quantum Chromodynamics (QCD 16),  4 july - 8 july 2016, Montpellier - FR}
 \author[label1]{Marco Frasca}
%  \cortext[cor0]{FAPESP CNPq-Brasil PhD student fellow.}
\ead{marcofrasca@mclink.it}
\address[label1]{Via Erasmo Gattamelata, 3
00176 Rome (Italy)}

\pagestyle{myheadings}
\markright{ }
\begin{abstract}
We derive the next-to-leading order correction to the Nambu-Jona-Lasinio model starting from quantum chromodynamics. So, we are able to fix the constants of the Nambu-Jona-Lasinio model from quantum chromodynamics and analyze the behavior of strong interactions at low energies. The technique is to expand in powers of currents the generating functional. We apply it to a simple Yukawa model with self-interaction showing how this has a Nambu-Jona-Lasinio model and its higher order corrections as a low-energy limit. The same is shown to happen for quantum chromodynamics in the chiral limit with two quarks. We prove that a consistent thermodynamic behaviour is obtained as expected for the given parameters.
\end{abstract}
% \begin{document}
\begin{keyword}  
%% keywords here, in the form: keyword \sep keyword
Nambu-Jona-Lasinio model \sep Non-local Nambu-Jona-Lasinio model \sep QCD \sep Finite temperature
%% MSC codes here, in the form: \MSC code \sep code
%% or \MSC[2008] code \sep code (2000 is the default)

\end{keyword}

\end{frontmatter}
%%%%%%%%%%%%
%\vspace*{-1.5cm}

Recent lattice studies for the gluon propagator \cite{Bogolubsky:2007ud,Cucchieri:2007md,Oliveira:2007px} and the spectrum \cite{Lucini:2004my,Chen:2005mg} showed evidence of a mass gap in a Yang-Mills theory without fermionic degrees of freedom. These results received theoretical support \cite{Cornwall:1981zr,Cornwall:2010bk,Dudal:2008sp,Tissier:2010ts,Tissier:2011ey,Frasca:2007uz,Frasca:2009yp} providing a closed form formula for the gluon propagator. An understanding of the gluon propagator is pivotal to derive the low-energy behaviour of QCD in a manageable effective theory. Some other results are also essential for this aim as the behaviour of the running coupling in the infared limit \cite{Nesterenko:1999np,Nesterenko:2001st,Nesterenko:2003xb,Baldicchi:2007ic,Baldicchi:2007zn,Bogolubsky:2009dc,Duarte:2016iko} (see also the review \cite{Deur:2016tte}) beside the gluon propagator. We will see that, for the latter, the instanton liquid plays an essential role \cite{Schafer:1996wv,Boucaud:2002fx}.

We have succeeded to show that a non-local Nambu-Jona-Lasinio model (nlNJL) represents the low energy limit of QCD \cite{Frasca:2008zp,Frasca:2013kka,Frasca:2012iv,Frasca:2012eq,Frasca:2011bd}. Here we generalizes this approach to get higher order corrections to the nlNJL verifying its consistency in a fully thermodynamic computation. This gives strong support to the already postulated extensions to the NJL model \cite{Kashiwa:2006rc,Osipov:2006ns,Hiller:2008nu,Gatto:2010qs,Gatto:2010pt}. As a historical aside, we show how the well-known Yukawa model recovers a nlNJL model. This was already shown in \cite{Frasca:2016utr} and does justice {\it a posteriori} to Yukawa's great insight.

So, in order to understand the technique, we consider the simplest Yukawa model: A scalar field interacting with a quark field. One has
\begin{equation}
      L_Y=\bar q(i\slashed\partial-g\phi)q+\frac{1}{2}(\partial\phi)^2-\frac{\lambda}{4}\phi^4.
\end{equation}
Then, the generating functional is
\begin{eqnarray}
      Z_Y[\bar\eta,\eta]\eqnsgn{=}\int [d\bar q][dq]\exp\left[-i\int d^4x\bar q\left(i\slashed\partial
			-g\frac{\delta}{i\delta j}\right)q\right]\times\qquad \nonumber \\
      &&\left.\exp(iW[j])\right|_{j=0}\exp\left[i\int d^4x\left(\bar\eta q+\bar q\eta\right)\right]
\end{eqnarray}
being
\begin{eqnarray}
\label{eq:Wj}
     W[j] \eqnsgn{=} W[0]+\int d^4xj(x)\phi_0(x) \\
		\eqnsgn{+}\frac{1}{2}\int d^4xd^4x_1j(x)\Delta(x-x_1)j(x_1)+O(j^3).  \nonumber
\end{eqnarray}
We have supposed to have exactly solved for the equation of motion of the scalar field obtaining the 1- and 2-point functions written as $\phi_0(x)$ and $\Delta(x-x_1)$ \cite{Frasca:2013tma,Frasca:2015wva}. The idea behind the functional (\ref{eq:Wj}) is a current expansion already devised in the '80 \cite{Cahill:1985mh}. The 1-point function is given by
\begin{equation}
      \phi_0(x) = \mu\left(2/\lambda\right)^\frac{1}{4}{\rm sn}(p\cdot x+\theta,-1)
\end{equation}
being sn the snoidal Jacobi function, $\mu$ and $\theta$ arbitrary integration constants. The 2-point function is
\begin{equation}
      \Delta(p)=\sum_{n=0}^\infty\frac{B_n}{p^2-m_n^2+i\epsilon}
\end{equation}
with
\begin{equation}
      B_n=(2n+1)^2\frac{\pi^3}{4K^3(-1)}\frac{e^{-(n+\frac{1}{2})\pi}}{1+e^{-(2n+1)\pi}}.
\end{equation}
A mass spectrum is obtained given by
\begin{equation}
      m_n=(2n+1)(\pi/2K(-1))\left(\lambda/2\right)^{\frac{1}{4}}\mu
\end{equation}
and $K(-1)\approx 1.3111028777$ is an elliptic integral. We can see that the action for the quark field, after the introduction of the n-point functions, recovers a nlNJL model and its possible higher order corrections
\begin{eqnarray}
      S_{q}\eqnsgn{=}\int d^4x\bar q\left(i\slashed\partial-g\phi_0(x)\right. \nonumber \\
			\eqnsgn{-}\left.\frac{g^2}{2}\int d^4x'\Delta(x-x')\bar q q\right)q+\ldots,
\end{eqnarray}
where dots imply higher powers of the quark current. If we average on the phase of the background field $\phi_0$, that is arbitrary, and fix this constant to zero, we are left with 
\begin{eqnarray}
      S_{NJL}\eqnsgn{=}\int d^4x\left[\bar q(x)i{\slashed\partial}q(x)\right. \\
			\eqnsgn{-}\left.\frac{g^2}{2}\int d^4x'\Delta(x-x')\bar q(x)q(x)\bar q(x')q(x')\right]. \nonumber
\end{eqnarray}
We note that, in the low-energy (local) limit $\Delta(x-x')=-(C_0/m_0^2)\delta^4(x-x')$, that leaves
\begin{eqnarray}
      S_{NJL}\eqnsgn{=}\int d^4x\left[\bar q(x)i{\slashed\partial}q(x)\right. \nonumber \\
			\eqnsgn{+}\left.C_0\frac{g^2}{2m_0^2}\bar q(x)q(x)\bar q(x)q(x)\right]
\end{eqnarray}
and the NJL constant $G=C_0g^2/m_0^2$ is fixed by the original Yukawa Lagrangian that describes the microscopic dynamics of the model.

\begin{figure}[t]
\centerline{\includegraphics[width=77.5mm]{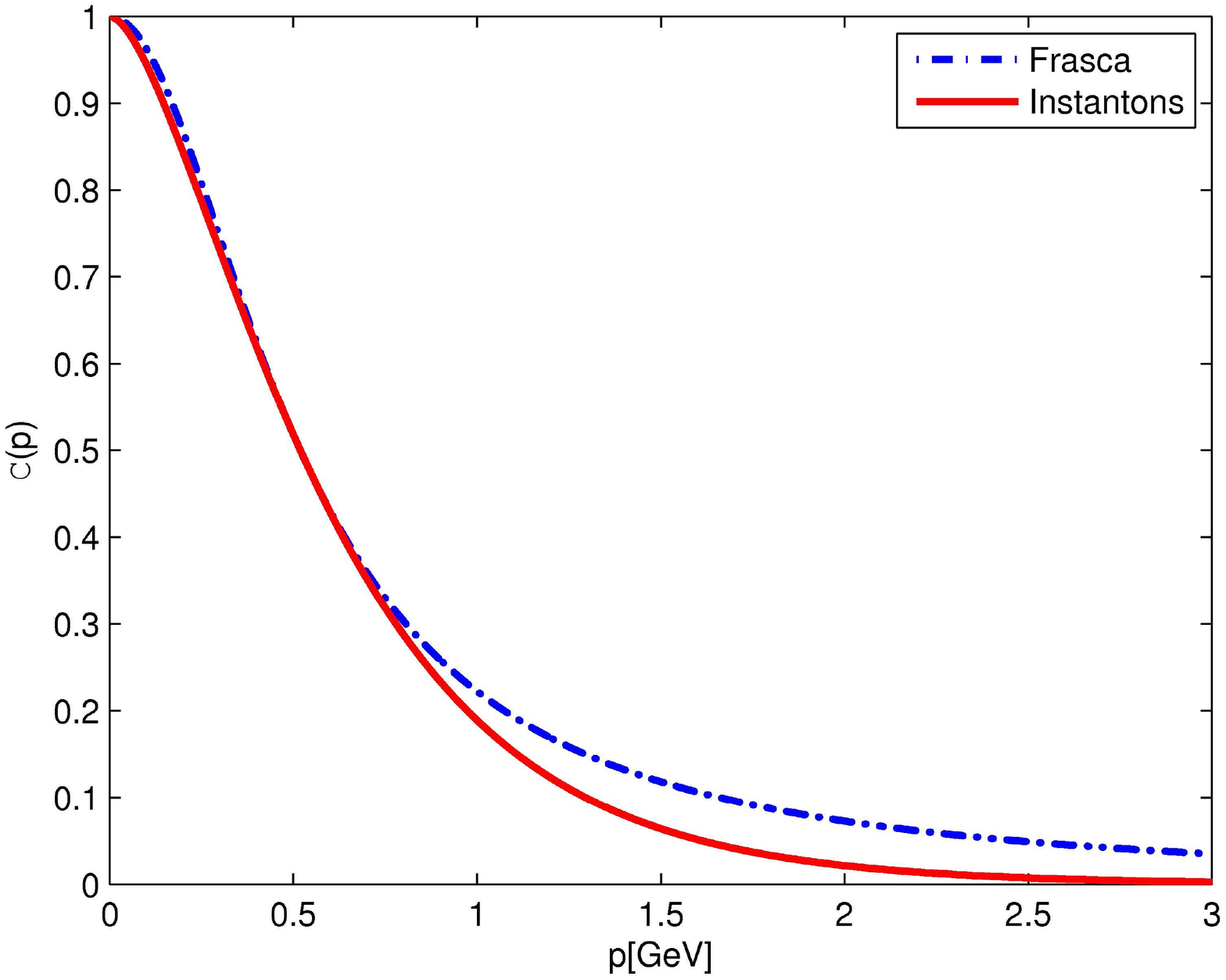}}
  \caption{Comparison of eq.(\ref{eq:kern}) with the one of an instanton liquid given in Ref~\cite{Schafer:1996wv}. \label{fig:fig1}}
\end{figure}

Yang-Mills theory admits a set of exact solutions and all the correlation functions can be obtained \cite{Frasca:2015yva}. Indeed, such solutions map on scalar field solutions \cite{Frasca:2009yp}. This means that the following functional series holds in this case 
\begin{eqnarray}
       W_{YM}[j,\epsilon,\bar\epsilon]\eqnsgn{=}\int d^4x\bar\epsilon(x)G(x)\epsilon(x)+\int d^4x\Phi_\mu^a(x)j^{\mu a}(x) \nonumber \\
				\eqnsgn{-}\int d^4x_1d^4x_2j^{\mu a}(x_1)\Delta_{\mu\nu}^{ab}(x_1-x_2)j^{\nu b}(x_2) \nonumber \\
	 \eqnsgn{+}\int d^4xd^4x_1d^4x_2d^4x_3\Phi^{\mu a}(x)\Delta_{\mu\nu}^{ab}(x-x_1)j^{\nu b}(x_1)\times \nonumber \\
	&&\Delta_{\kappa\lambda}^{cd}(x-x_2)j^{\kappa c}(x_2)\Delta^{\lambda de}_\rho(x-x_3)j^{\rho e}(x_3) \nonumber \\
	\eqnsgn{+} \int d^4xd^4x_1d^4x_2d^4x_3d^4x_4\Delta_{\mu\nu}^{ab}(x-x_1)j^{\mu a}(x_1)\times \nonumber \\
	&&\Delta_{\lambda}^{\nu bc}(x-x_2)j^{\lambda c}(x_2)\times \nonumber \\
	&&\Delta^{de}_{\kappa\rho}(x-x_3)j^{\kappa d}(x_3)\times \nonumber \\
	&&\Delta^{\rho ef}_{\theta}(x-x_4)j^{\theta f}(x_4) + O(j^5).	
\end{eqnarray}
Here $G(x)$ is the ghost two-point function, $\Phi^{\mu a}(x)$ is the Yang-Mills 1-point function and $\Delta_{\mu\nu}^{ab}(x_1-x_2)$ the Yang-Mills 2-point function. Using the substitution with quark currents $j^{\mu a}(x_1)\rightarrow \sum_q\bar q(x)\frac{\lambda^a}{2}\gamma^\mu q(x)$ we get a nlNJL model from QCD at low-energies 
\begin{eqnarray}
      S_{NJL}\eqnsgn{=}-g\int d^4x\Phi_\mu^a(x)\sum_q\bar q(x)\frac{\lambda^a}{2}\gamma^\mu q(x) \nonumber \\
	\eqnsgn{-}g^2\int d^4x_1d^4x_2\sum_q\bar q(x_1)\frac{\lambda^a}{2}\gamma^\mu q(x_1)\Delta(x_1-x_2)\times \nonumber \\
	&&\sum_{q'}\bar q'(x_2)\frac{\lambda^a}{2}\gamma_\mu q'(x_2) \nonumber \\
	\eqnsgn{-}g^3\int d^4xd^4x_1d^4x_2d^4x_3\Phi^{\mu a}(x)\Delta(x-x_1)\times \nonumber \\
	&&\sum_q\bar q(x_1)\frac{\lambda^a}{2}\gamma_\mu q(x_1)\times \nonumber \\
	&&\Delta(x-x_2)\sum_{q'}\bar q'(x_2)\frac{\lambda^b}{2}\gamma^\nu q'(x_2)\Delta(x-x_3)\times \nonumber \\
	&&\sum_{q''}\bar q''(x_3)\frac{\lambda^b}{2}\gamma_\nu q''(x_3) \nonumber \\
      \eqnsgn{+}g^4\int d^4xd^4x_1d^4x_2d^4x_3d^4x_4\Delta(x-x_1)\times \nonumber \\
			&&\sum_q\bar q(x_1)\frac{\lambda^a}{2}\gamma^\mu q(x_1)\times \nonumber \\
	&&\Delta(x-x_2)\sum_{q'}\bar q'(x_2)\frac{\lambda^a}{2}\gamma_\mu q'(x_2)\times \nonumber \\
      &&\Delta(x-x_3)\sum_{q''}\bar q''(x_3)\frac{\lambda^b}{2}\gamma^\nu q''(x_3)\times \nonumber \\
	&&\Delta(x-x_4)\sum_{q'''}\bar q'''(x_4)\frac{\lambda^b}{2}\gamma_\nu q'''(x_4).	
\end{eqnarray}
For the non-local limit the propagator yields 
\begin{eqnarray}
\label{eq:kern}
    {\cal G}(p)\eqnsgn{=}-\frac{1}{2}g^2\Delta(p) \\
		\eqnsgn{=}-\frac{1}{2}g^2\sum_{n=0}^\infty\frac{B_n}{p^2-(2n+1)^2(\pi/2K(i))^2\tilde\sigma+i\epsilon}
      =\frac{G}{2}{\cal C}(p)  \nonumber
\end{eqnarray}
to be compared with \cite{Schafer:1996wv} for instanton liquid at $\sqrt{\sigma}=0.417\ GeV$ and $d^{-1}=0.58\ GeV$ 
\begin{equation}
      \mathcal{C}_I(\xi)=4\pi^2 d^2\left\{\xi\frac{d}{d\xi}\big[I_0(\xi)K_0(\xi)-I_1(\xi)K_1(\xi)\big]\right\}^2
\end{equation}
with $\xi=\frac{|p| d}{2}$. The comparison is given in fig.~\ref{fig:fig1} and we get a strikingly good agreement.

From the nlNJL model we take the local limit and two-flavor approximation that are enough for our aims. So, for $\psi=(u,d)$ and averaging on the phase of $\Phi^{\mu a}(x)$ as already done for the Yukawa model, we get
\begin{eqnarray}
  W_{NJL}[q,\bar q] \eqnsgn{=} \frac{G}{2}\int d^4x\left[\bar\psi(x)\gamma^\mu\psi(x)\bar\psi(x)\gamma_\mu\psi(x)\right. \nonumber \\
	\eqnsgn{+}\left.\bar\psi(x)\gamma^\mu{\bm\tau}\psi(x)\bar\psi(x)\gamma_\mu{\bm\tau}\psi(x)\right] \nonumber \\
	\eqnsgn{+}G_8\int d^4x\left[\bar\psi(x)\gamma^\mu\psi(x)\bar\psi(x)\gamma_\mu\psi(x)\right. \nonumber \\
	\eqnsgn{+}\left.\bar\psi(x)\gamma^\mu{\bm\tau}\psi(x)\bar\psi(x)\gamma_\mu{\bm\tau}\psi(x)\right]\times \nonumber \\
	&& \left[\bar\psi(x)\gamma^\mu\psi(x)\bar\psi(x)\gamma_\mu\psi(x)\right. \nonumber \\
	\eqnsgn{+}\left.\bar\psi(x)\gamma^\mu{\bm\tau}\psi(x)\bar\psi(x)\gamma_\mu{\bm\tau}\psi(x)\right]
\end{eqnarray}
with $G$ the usual NJL coupling given by eq.(\ref{eq:kern}) and we have
\begin{eqnarray} 
    G_8 \eqnsgn{=} 4g^{-4}\left[{\cal G}(0)\right]^4=\frac{g^4}{4(\pi/2K(-1))^8{\tilde\sigma}^4}\left[\sum_{n=0}^{\infty}\frac{B_n}{(2n+1)^2}\right]^4 \nonumber \\
		 \eqnsgn{\approx}0.096\frac{g^4}{{\tilde\sigma}^4}.
\end{eqnarray}
This shows that the next-to-leading order correction to the NJL model is an eight quark interaction term in agreement with what recently postulated \cite{Kashiwa:2006rc,Osipov:2006ns,Hiller:2008nu,Gatto:2010qs,Gatto:2010pt}. The question we want to answer is if such a term we derived from QCD is consistent with the expected thermodynamic behaviour of the theory. We are going to discuss this point below.

To get an understanding of this model we have to analyze the gap equation. We have to solve the self-consistent set of equations \cite{Hiller:2008nu}, in the chiral limit $m_u=m_d=0$,
\begin{eqnarray}
    M+Gh+\frac{3}{16}G_8h^3\eqnsgn{=}0 \nonumber \\
    h(M)+\frac{NN_f}{2\pi^2}MJ_0(M^2)\eqnsgn{=}0.
\end{eqnarray}
where $J_0(M^2)$ is the first of NJL integrals and is yielded, at zero temperature and chemical potential, by
\begin{eqnarray}
J_0(M^2)\eqnsgn{=}16\pi^2i\int_\Lambda\frac{d^4p}{(2\pi)^4}\frac{1}{p^2-M^2} \nonumber \\
\eqnsgn{=}\Lambda^2-M^2\ln\left(1+\frac{M^2}{\Lambda^2}\right).
\end{eqnarray}
A cut-off $\Lambda$ is needed to regularize the theory. We get $M_{eff}=0.63\ GeV$ for the quark effective mass with a cut-off $\Lambda=0.747\ GeV$, assuming $G\approx 9.37\ GeV^{-2}$ (see \cite{Frasca:2012eq}) and $G_8=G^4/(\sqrt{2}g)^4$.

\begin{figure}[t]
\centerline{\includegraphics[width=77.5mm]{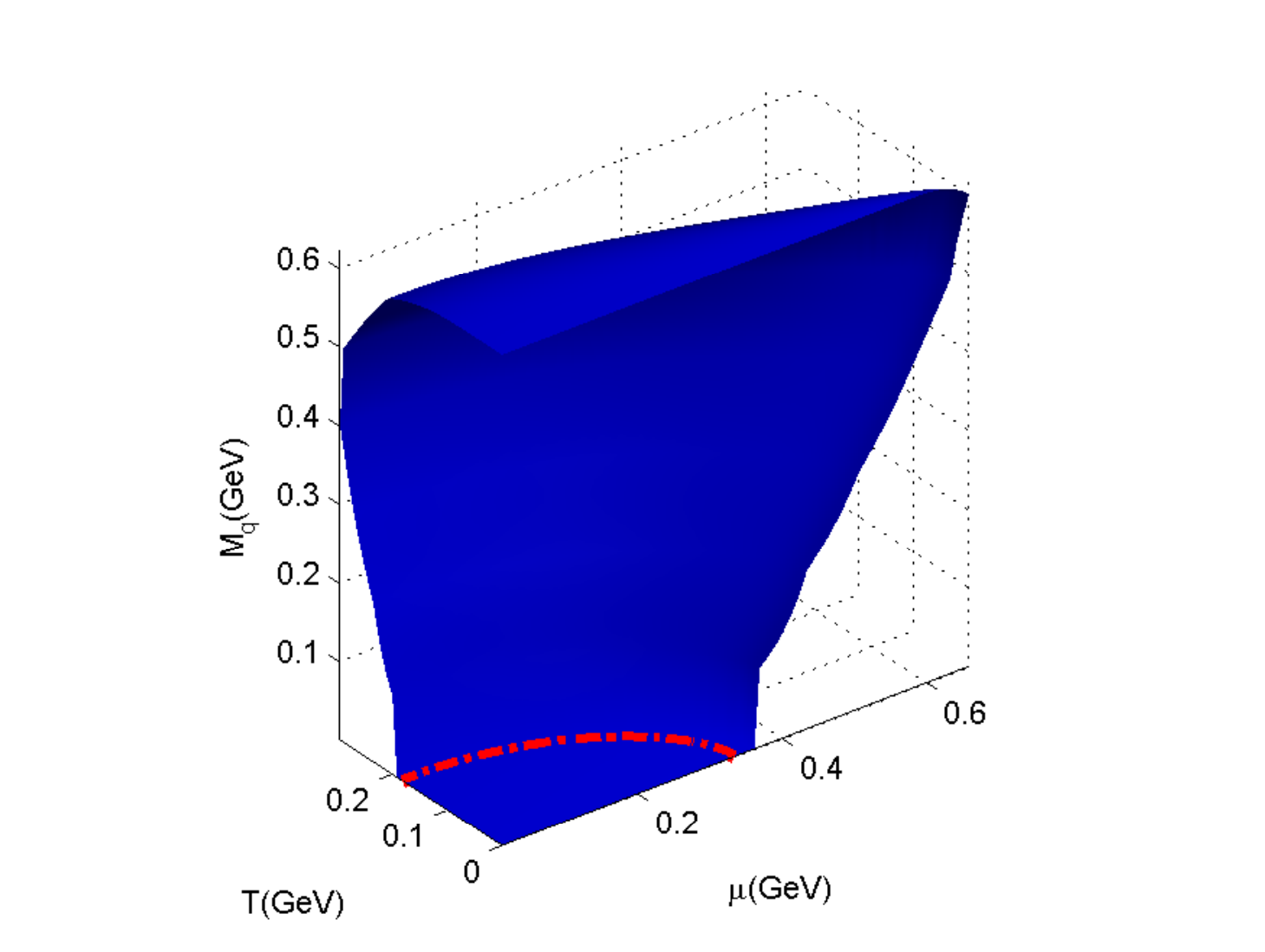}}
  \caption{Effective quark mass as a function of $\mu$ and $T$. The critical line $T_c=T_c(\mu)$ is plotted in dot-dashed red.\label{fig:fig2}}
\end{figure}

At finite temperature and chemical potential one has for the gap equation
\begin{equation}
    h(M)+\frac{NN_f}{2\pi^2}MJ_0(M^2,T,\mu)=0.
\end{equation}
with
\begin{eqnarray}
J_0(M^2,T,\mu)\eqnsgn{=}J_0(M^2)-4\int_0^\Lambda dp\frac{p^2}{E_p}\times \nonumber \\
&&\left(\frac{1}{1+e^{\beta(E_p-\mu)}}+\frac{1}{1+e^{\beta(E_p+\mu)}}\right)
\end{eqnarray}
being $E_p=\sqrt{p^2+M^2}$. We can treat this integral numerically to obtain the phase diagram in this case. But we note a simple fact, already at this stage, in the chiral limit $M\rightarrow 0$ the critical temperature is left untouched by the eight quark term. This can be understood by noticing that
\begin{equation}
     \lim_{M\rightarrow 0}\frac{h(M)}{M}=-\frac{1}{G}
\end{equation}
leaving us with the usual gap equation to determine the critical temperature \cite{GomezDumm:2004sr,Frasca:2011bd}.
Fig.~\ref{fig:fig2} presents the quark condensate $M_q(T,\mu)=-G\langle \bar qq\rangle$ and the critical line $T_c=T_c(\mu)$ at $M_q(T_c,\mu)=0$ with the given parameters

We have shown how to derive a nlNJL model starting from a Yukawa theory. Recent studies permit to obtain the NJL model directly from QCD in the low-energy limit with the same technique. We have found the next-to-leading order correction to the NJL model yielding and 8-quark interaction term as already postulated in literature. The critical temperature, in the chiral limit setting quark masses to zero, is unaffected by this correction. A consistent thermodynamic behaviour is also obtained as expected for the given parameters, particularly, we recover a consistent curve for the critical temperature as a function of the chemical potential. Further studies will be needed to understand the physical spectrum of the theory in the low-energy limit. In this respect, it is worthwhile to point out that a non-confining theory, as NJL is, yields just bound states and no free quarks. Low-energy physical states cannot coincide with those in the ultraviolet limit.

%% Acknowledgements %%
I have to thank Silvio Sorella for several enlightening discussions during the Conference.


\begin{thebibliography}{999}

%%%% Lattice data %%%%%%%%
  
%\cite{Bogolubsky:2007ud}
\bibitem{Bogolubsky:2007ud}
  I.~L.~Bogolubsky, E.~M.~Ilgenfritz, M.~Muller-Preussker, A.~Sternbeck,
  %``The Landau gauge gluon and ghost propagators in 4D SU(3) gluodynamics in large lattice volumes,''
  PoS LAT2007, 290 (2007).
  %[arXiv:0710.1968 [hep-lat]].
  
%\cite{Cucchieri:2007md}
\bibitem{Cucchieri:2007md}
  A.~Cucchieri, T.~Mendes,
  %``What's up with IR gluon and ghost propagators in Landau gauge? A puzzling answer from huge lattices,''
  PoS LAT2007, 297 (2007).
  %[arXiv:0710.0412 [hep-lat]]. 
   
%\cite{Oliveira:2007px}
\bibitem{Oliveira:2007px}
  O.~Oliveira, P.~J.~Silva, E.~M.~Ilgenfritz, A.~Sternbeck,
  %``The Gluon propagator from large asymmetric lattices,''
  PoS LAT2007, 323 (2007).
  %[arXiv:0710.1424 [hep-lat]].
	
%\cite{Lucini:2004my}
\bibitem{Lucini:2004my} 
  B.~Lucini, M.~Teper and U.~Wenger,
  %``Glueballs and k-strings in SU(N) gauge theories: Calculations with improved operators,''
  JHEP {\bf 0406}, 012 (2004).
  %doi:10.1088/1126-6708/2004/06/012
  %[hep-lat/0404008].
  %%CITATION = doi:10.1088/1126-6708/2004/06/012;%%
  %145 citations counted in INSPIRE as of 14 févr. 2016
	
%\cite{Chen:2005mg}
\bibitem{Chen:2005mg} 
  Y.~Chen {\it et al.},
  %``Glueball spectrum and matrix elements on anisotropic lattices,''
  Phys.\ Rev.\ D {\bf 73}, 014516 (2006).
  %doi:10.1103/PhysRevD.73.014516
  %[hep-lat/0510074].
  %%CITATION = doi:10.1103/PhysRevD.73.014516;%%
  %311 citations counted in INSPIRE as of 14 févr. 2016

%%%%% Theoretical understanding of Yang-Mills thoery %%%%%%%%%%

%\cite{Cornwall:1981zr}
\bibitem{Cornwall:1981zr} 
  J.~M.~Cornwall,
  %``Dynamical Mass Generation in Continuum QCD,''
  Phys.\ Rev.\ D {\bf 26}, 1453 (1982).
  %%CITATION = PHRVA,D26,1453;%%
  %806 citations counted in INSPIRE as of 04 Jun 2015
	
\bibitem{Cornwall:2010bk}	
	J.~M.~Cornwall, J.~Papavassiliou, D.~Binosi, 
  {\sl The Pinch Technique and its Applications to Non-Abelian Gauge Theories}, 	
	(Cambridge University Press, Cambridge, 2010).	

%\cite{Dudal:2008sp}
\bibitem{Dudal:2008sp} 
  D.~Dudal, J.~A.~Gracey, S.~P.~Sorella, N.~Vandersickel and H.~Verschelde,
  %``A Refinement of the Gribov-Zwanziger approach in the Landau gauge: Infrared propagators in harmony with the lattice results,''
  Phys.\ Rev.\ D {\bf 78}, 065047 (2008)
  [arXiv:0806.4348 [hep-th]].
  %%CITATION = ARXIV:0806.4348;%%
  %283 citations counted in INSPIRE as of 04 Jun 2015
	
%\cite{Tissier:2010ts}
\bibitem{Tissier:2010ts} 
  M.~Tissier and N.~Wschebor,
  %``Infrared propagators of Yang-Mills theory from perturbation theory,''
  Phys.\ Rev.\ D {\bf 82}, 101701 (2010)
  [arXiv:1004.1607 [hep-ph]].
  %%CITATION = ARXIV:1004.1607;%%
  %44 citations counted in INSPIRE as of 04 Jun 2015
		
%\cite{Tissier:2011ey}
\bibitem{Tissier:2011ey} 
  M.~Tissier and N.~Wschebor,
  %``An Infrared Safe perturbative approach to Yang-Mills correlators,''
  Phys.\ Rev.\ D {\bf 84}, 045018 (2011)
  [arXiv:1105.2475 [hep-th]].
  %%CITATION = ARXIV:1105.2475;%%
  %35 citations counted in INSPIRE as of 04 Jun 2015
	  
%\cite{Frasca:2007uz}
\bibitem{Frasca:2007uz} 
  M.~Frasca,
  %``Infrared Gluon and Ghost Propagators,''
  Phys.\ Lett.\ B {\bf 670}, 73 (2008).
  %[arXiv:0709.2042 [hep-th]].
  %%CITATION = ARXIV:0709.2042;%%
	  
%\cite{Frasca:2009yp}
\bibitem{Frasca:2009yp}
  M.~Frasca,
  %``Mapping a Massless Scalar Field Theory on a Yang-Mills Theory: Classical Case,''
  Mod.\ Phys.\ Lett.\  {\bf A24}, 2425-2432 (2009).
  %[arXiv:0903.2357 [math-ph]].
	
%%%% Nesterenko %%%%%%

%\cite{Nesterenko:1999np}
\bibitem{Nesterenko:1999np} 
  A.~V.~Nesterenko,
  %``Quark - anti-quark potential in the analytic approach to QCD,''
  Phys.\ Rev.\ D {\bf 62}, 094028 (2000),
  %doi:10.1103/PhysRevD.62.094028
  [hep-ph/9912351].
  %%CITATION = doi:10.1103/PhysRevD.62.094028;%%
  %76 citations counted in INSPIRE as of 17 Aug 2016

%\cite{Nesterenko:2001st}
\bibitem{Nesterenko:2001st} 
  A.~V.~Nesterenko,
  %``New analytic running coupling in space - like and time - like regions,''
  Phys.\ Rev.\ D {\bf 64}, 116009 (2001),
  %doi:10.1103/PhysRevD.64.116009
  [hep-ph/0102124].
  %%CITATION = doi:10.1103/PhysRevD.64.116009;%%
  %63 citations counted in INSPIRE as of 17 Aug 2016
	
%\cite{Nesterenko:2003xb}
\bibitem{Nesterenko:2003xb} 
  A.~V.~Nesterenko,
  %``Analytic invariant charge in QCD,''
  Int.\ J.\ Mod.\ Phys.\ A {\bf 18}, 5475 (2003),
  %doi:10.1142/S0217751X0301704X
  [hep-ph/0308288].
  %%CITATION = doi:10.1142/S0217751X0301704X;%%
  %79 citations counted in INSPIRE as of 17 Aug 2016

%\cite{Baldicchi:2007ic}
\bibitem{Baldicchi:2007ic} 
  M.~Baldicchi, A.~V.~Nesterenko, G.~M.~Prosperi, D.~V.~Shirkov and C.~Simolo,
  %``Bound state approach to the QCD coupling at low energy scales,''
  Phys.\ Rev.\ Lett.\  {\bf 99}, 242001 (2007),
  %doi:10.1103/PhysRevLett.99.242001
  [arXiv:0705.0329 [hep-ph]].
  %%CITATION = doi:10.1103/PhysRevLett.99.242001;%%
  %47 citations counted in INSPIRE as of 17 Aug 2016

%\cite{Baldicchi:2007zn}
\bibitem{Baldicchi:2007zn} 
  M.~Baldicchi, A.~V.~Nesterenko, G.~M.~Prosperi and C.~Simolo,
  %``QCD coupling below 1 GeV from quarkonium spectrum,''
  Phys.\ Rev.\ D {\bf 77}, 034013 (2008),
  %doi:10.1103/PhysRevD.77.034013
  [arXiv:0705.1695 [hep-ph]].
  %%CITATION = doi:10.1103/PhysRevD.77.034013;%%
  %39 citations counted in INSPIRE as of 17 Aug 2016
	
%%%% Running coupling on lattice %%%%%%
	
%\cite{Bogolubsky:2009dc}
\bibitem{Bogolubsky:2009dc}
  I.~L.~Bogolubsky, E.~M.~Ilgenfritz, M.~Muller-Preussker, A.~Sternbeck,
  %``Lattice gluodynamics computation of Landau gauge Green's functions in the deep infrared,''
  Phys.\ Lett.\  {\bf B676}, 69-73 (2009),
  [arXiv:0901.0736 [hep-lat]].
	
%\cite{Duarte:2016iko}
\bibitem{Duarte:2016iko} 
  A.~G.~Duarte, O.~Oliveira and P.~J.~Silva,
  %``Lattice Gluon and Ghost Propagators, and the Strong Coupling in Pure SU(3) Yang-Mills Theory: Finite Lattice Spacing and Volume Effects,''
  Phys.\ Rev.\ D {\bf 94}, no. 1, 014502 (2016),
  %doi:10.1103/PhysRevD.94.014502
  [arXiv:1605.00594 [hep-lat]].
  %%CITATION = doi:10.1103/PhysRevD.94.014502;%%
  %5 citations counted in INSPIRE as of 17 Aug 2016
	
%\cite{Deur:2016tte}
\bibitem{Deur:2016tte} 
  A.~Deur, S.~J.~Brodsky and G.~F.~de Teramond,
  %``The QCD Running Coupling,''
  Prog.\ Part.\ Nucl.\ Phys.\  {\bf 90}, 1 (2016),
  %doi:10.1016/j.ppnp.2016.04.003
  [arXiv:1604.08082 [hep-ph]].
  %%CITATION = doi:10.1016/j.ppnp.2016.04.003;%%
  %9 citations counted in INSPIRE as of 24 Aug 2016
	
%%%% Instanton liquid %%%%
    
%\cite{Schafer:1996wv}
\bibitem{Schafer:1996wv}
  T.~Sch\"afer and E.~V.~Shuryak,
  %``Instantons in QCD,''
  Rev.\ Mod.\ Phys.\  {\bf 70}, 323 (1998)
  %[arXiv:hep-ph/9610451].
  %%CITATION = RMPHA,70,323;%%
	
%\cite{Boucaud:2002fx}
\bibitem{Boucaud:2002fx} 
  P.~Boucaud, F.~De Soto, A.~Le Yaouanc, J.~P.~Leroy, J.~Micheli, H.~Moutarde, O.~Pene and J.~Rodriguez-Quintero,
  %``The Strong coupling constant at small momentum as an instanton detector,''
  JHEP {\bf 0304}, 005 (2003),
  %doi:10.1088/1126-6708/2003/04/005
  [hep-ph/0212192].
  %%CITATION = doi:10.1088/1126-6708/2003/04/005;%%
  %46 citations counted in INSPIRE as of 17 Aug 2016
	
%%%% nlNJL as low-energy limit of QCD %%%%

%\cite{Frasca:2008zp}
\bibitem{Frasca:2008zp} 
  M.~Frasca,
  %``Infrared QCD,''  
  Int.\ J.\ Mod.\ Phys.\ E {\bf 18}, 693 (2009),  
  [arXiv:0803.0319 [hep-th]].  
  %%CITATION = ARXIV:0803.0319;%%  
  %14 citations counted in INSPIRE as of 16 Sep 2013
		
%\cite{Frasca:2013kka}
\bibitem{Frasca:2013kka} 
  M.~Frasca,
  %``$\rho$ condensation and physical parameters,''
  JHEP {\bf 1311}, 099 (2013),
  %doi:10.1007/JHEP11(2013)099
  [arXiv:1309.3966 [hep-ph]].
  %%CITATION = doi:10.1007/JHEP11(2013)099;%%
  %10 citations counted in INSPIRE as of 14 Feb 2016
  
%\cite{Frasca:2012iv}
\bibitem{Frasca:2012iv} 
  M.~Frasca,
  %``Low energy limit of QCD and the emerging of confinement,''
  Nucl.\ Phys.\ Proc.\ Suppl.\  {\bf 234}, 329 (2013),
  [arXiv:1208.3756 [hep-ph]].
  %%CITATION = ARXIV:1208.3756;%%
  %1 citations counted in INSPIRE as of 28 Aug 2013
  
%\cite{Frasca:2012eq}
\bibitem{Frasca:2012eq} 
  M.~Frasca,
  %``Low-energy QCD from first principles,''
  AIP Conf.\ Proc.\  {\bf 1492}, 177 (2012),
  [arXiv:1208.0486 [hep-ph]].
  %%CITATION = ARXIV:1208.0486;%%
  %1 citations counted in INSPIRE as of 28 Aug 2013

%\cite{Frasca:2011bd}
\bibitem{Frasca:2011bd} 
  M.~Frasca,
  %``Chiral symmetry in the low-energy limit of QCD at finite temperature,''
  Phys.\ Rev.\ C {\bf 84}, 055208 (2011),
  [arXiv:1105.5274 [hep-ph]].
  %%CITATION = ARXIV:1105.5274;%%
  %9 citations counted in INSPIRE as of 28 Aug 2013
	
%%%% Eight quark term %%%%
	
%\cite{Kashiwa:2006rc}
\bibitem{Kashiwa:2006rc} 
  K.~Kashiwa, H.~Kouno, T.~Sakaguchi, M.~Matsuzaki and M.~Yahiro,
  %``Chiral phase transition in an extended NJL model with higher-order multi-quark interactions,''
  Phys.\ Lett.\ B {\bf 647}, 446 (2007),
  %doi:10.1016/j.physletb.2007.01.061
  [nucl-th/0608078].
  %%CITATION = doi:10.1016/j.physletb.2007.01.061;%%
  %55 citations counted in INSPIRE as of 22 Apr 2016
	
%\cite{Osipov:2006ns}
\bibitem{Osipov:2006ns} 
  A.~A.~Osipov, B.~Hiller, A.~H.~Blin and J.~da Providencia,
  %``Effects of eight-quark interactions on the hadronic vacuum and mass spectra of light mesons,''
  Annals Phys.\  {\bf 322}, 2021 (2007),
  %doi:10.1016/j.aop.2006.08.004
  [hep-ph/0607066].
  %%CITATION = doi:10.1016/j.aop.2006.08.004;%%
  %52 citations counted in INSPIRE as of 24 Mar 2016
	
%\cite{Hiller:2008nu}
\bibitem{Hiller:2008nu} 
  B.~Hiller, J.~Moreira, A.~A.~Osipov and A.~H.~Blin,
  %``The Phase diagram for the Nambu-Jona-Lasinio model with 't Hooft and eight-quark interactions,''
  Phys.\ Rev.\ D {\bf 81}, 116005 (2010),
  %doi:10.1103/PhysRevD.81.116005
  [arXiv:0812.1532 [hep-ph]].
  %%CITATION = doi:10.1103/PhysRevD.81.116005;%%
  %35 citations counted in INSPIRE as of 23 Mar 2016

%\cite{Gatto:2010qs}
\bibitem{Gatto:2010qs} 
  R.~Gatto and M.~Ruggieri,
  %``Dressed Polyakov loop and phase diagram of hot quark matter under magnetic field,''
  Phys.\ Rev.\ D {\bf 82}, 054027 (2010),
  %doi:10.1103/PhysRevD.82.054027
  [arXiv:1007.0790 [hep-ph]].
  %%CITATION = doi:10.1103/PhysRevD.82.054027;%%
  %121 citations counted in INSPIRE as of 22 Apr 2016
	
%\cite{Gatto:2010pt}
\bibitem{Gatto:2010pt} 
  R.~Gatto and M.~Ruggieri,
  %``Deconfinement and Chiral Symmetry Restoration in a Strong Magnetic Background,''
  Phys.\ Rev.\ D {\bf 83}, 034016 (2011),
  %doi:10.1103/PhysRevD.83.034016
  [arXiv:1012.1291 [hep-ph]].
  %%CITATION = doi:10.1103/PhysRevD.83.034016;%%
  %146 citations counted in INSPIRE as of 22 Apr 2016

%%%% Full paper %%%%
	
%\cite{Frasca:2016utr}
\bibitem{Frasca:2016utr} 
  M.~Frasca,
  %``Correction to Nambu-Jona-Lasinio model from QCD at the next-to-leading order,''
  arXiv:1604.06640 [hep-ph].
  %%CITATION = ARXIV:1604.06640;%%
	
%%%% Scalar Field %%%%

%\cite{Frasca:2013tma}
\bibitem{Frasca:2013tma} 
  M.~Frasca,
  %``Scalar field theory in the strong self-interaction limit,''
  Eur.\ Phys.\ J.\ C {\bf 74}, 2929 (2014),
  %doi:10.1140/epjc/s10052-014-2929-9
  [arXiv:1306.6530 [hep-ph]].
  %%CITATION = doi:10.1140/epjc/s10052-014-2929-9;%%
  %10 citations counted in INSPIRE as of 17 Aug 2016
	
%\cite{Frasca:2015wva}
\bibitem{Frasca:2015wva} 
  M.~Frasca,
  %``A theorem on the Higgs sector of the Standard Model,''
  Eur.\ Phys.\ J.\ Plus {\bf 131}, no. 6, 199 (2016),
  %doi:10.1140/epjp/i2016-16199-x
  [arXiv:1504.02299 [hep-ph]].
  %%CITATION = doi:10.1140/epjp/i2016-16199-x;%%
  %1 citations counted in INSPIRE as of 17 Aug 2016
	
%%%% Current expansion %%%%
  
%\cite{Cahill:1985mh}
\bibitem{Cahill:1985mh}
  R.~T.~Cahill and C.~D.~Roberts,
  %``Soliton Bag Models of Hadrons from QCD,''
  Phys.\ Rev.\  D {\bf 32}, 2419 (1985).
  %%CITATION = PHRVA,D32,2419;%%
	
%%%%% My exact solution %%%%

%\cite{Frasca:2015yva}
\bibitem{Frasca:2015yva} 
  M.~Frasca,
  %``Quantum Yang-Mills field theory,''
  arXiv:1509.05292 [math-ph].
  %%CITATION = ARXIV:1509.05292;%%
  %2 citations counted in INSPIRE as of 17 Aug 2016
	
%%%% My numbers for NJL %%%%

%\cite{Frasca:2012eq}
%\bibitem{Frasca:2012eq} 
%  M.~Frasca,
%  %``Low-energy QCD from first principles,''
%  AIP Conf.\ Proc.\  {\bf 1492}, 177 (2012),
%  [arXiv:1208.0486 [hep-ph]].
%  %%CITATION = ARXIV:1208.0486;%%
%  %1 citations counted in INSPIRE as of 28 Aug 2013

%%%% Scoccola %%%%
	
%\cite{GomezDumm:2004sr}
\bibitem{GomezDumm:2004sr} 
  D.~Gomez Dumm and N.~N.~Scoccola,
  %``Characteristics of the chiral phase transition in nonlocal quark models,''
  Phys.\ Rev.\ C {\bf 72}, 014909 (2005),
  %doi:10.1103/PhysRevC.72.014909
  [hep-ph/0410262].
  %%CITATION = doi:10.1103/PhysRevC.72.014909;%%
  %30 citations counted in INSPIRE as of 14 févr. 2016
	
\end{thebibliography}
\end{document}